\documentstyle[aps,prb]{revtex}

        \def\LaTwelve{La$_{1.88}$Sr$_{0.12}$CuO$_4$ }
        \def\LaOxy{La$_2$CuO$_{4 + y}$ }
        \def\LaSr{La$_{2-x}$Sr$_x$CuO$_4$ }
        \def\LaPure{La$_2$CuO$_4$ }

        \def\LaTwelvens{La$_{1.88}$Sr$_{0.12}$CuO$_4$}  
        \def\LaOxyns{La$_2$CuO$_{4 + y}$}               
        \def\LaPurens{La$_2$CuO$_4$}                    

        \def\tc{$T_c$ }
        \def\tcns{$T_c$}                        
        \def\et{{\it et al. }}

\begin{document}
\input epsf.sty
\twocolumn[\hsize\textwidth\columnwidth\hsize\csname %
@twocolumnfalse\endcsname
\draft
\widetext


\title{Incommensurate Geometry of the Elastic Magnetic Peaks in
  Superconducting La$_{1.88}$Sr$_{0.12}$CuO$_4$}

\author{H. Kimura$^*$, H. Matsushita, K. Hirota, and Y. Endoh$^{\dag}$}
\address{Department of Physics, Tohoku University, Aramaki, Aoba-ku,
  Sendai 980-8578, Japan} 

\author{K. Yamada}
\address{Institute for Chemical Research, Kyoto University, Gokasho,
  Uji 610-0011, Japan}

\author{G. Shirane}
\address{Department of Physics, Brookhaven National Laboratory, Upton,
New York 11973}

\author{Y. S. Lee$^{\ddag}$, M. A. Kastner, and R. J. Birgeneau}
\address{Department of Physics and Center for Materials Science and
  Engineering, Massachusetts Institute of Technology, Cambridge,
  Massachusetts 02139}

\date{\today}
\maketitle

\begin{abstract}

  We report magnetic neutron scattering measurements of incommensurate
  magnetic order in a superconducting single crystal of \LaTwelvens.
  We find that the incommensurate wavevectors which describe the
  static magnetism do not lie along high-symmetry directions of the
  underlying CuO$_2$ lattice.  The positions of the elastic magnetic
  peaks are consistent with those found in excess-oxygen doped
  \LaOxyns.  This behavior differs from the precise magnetic order
  found in the low temperature tetragonal
  La$_{1.6-x}$Nd$_{0.4}$Sr$_x$CuO$_4$ material for which stripes of
  spin and charge have been observed.  These observations have clear
  implications for any stripe model proposed to describe the static
  magnetism in orthorhombic \LaPurens-based superconductors.

\end{abstract}

\


\phantom{.}
]
\narrowtext

In the lamellar copper-oxides, neutron scattering experiments have
shown that antiferromagnetic spin correlations are intimately
intertwined with the superconductivity
\cite{Birgeneau88,Yoshizawa88,Cheong91,Yamada98,Mook98}.  Current evidence indicates
that static incommensurate magnetism coexists with superconductivity,
especially for incommensurabilities near $\frac{1}{8}$ reciprocal
lattice units.  This was first observed by Tranquada and coworkers in
La$_{1.6-x}$Nd$_{0.4}$Sr$_x$CuO$_4$ samples\cite{Tranquada96,Tranquada97} 
which are tetragonal at low temperatures.  Subsequently, Suzuki \et  
\cite{Suzuki98} and  Kimura \et \cite{Kimura99} have observed
spin density wave (SDW) order in \LaTwelvens, and Lee \et \cite{Lee99} have
found similar SDW order in stage-4 \LaOxyns.  Note that these crystals
remain orthorhombic at the lowest measured temperatures.  Interestingly, 
for these latter two systems, the ordering temperature for the magnetism 
coincides with the superconducting transition temperature.

In \LaTwelvens, a few important features require further
clarification.  The detailed geometry of the SDW modulation needs to
be established.  For example, one must check to see if the SDW peaks
have modulation wavevectors which are like
La$_{1.6-x}$Nd$_{0.4}$Sr$_x$CuO$_4$ or, rather, if they more closely
resemble those in \LaOxyns.  In La$_{1.6-x}$Nd$_{0.4}$Sr$_x$CuO$_4$, a
quartet of SDW peaks are observed in a square-shaped arrangement
around the antiferromagnetic Bragg position\cite{Tranquada96,Tranquada97}.  
In this case, the
incommensurate wavevectors are aligned with the Cu-O-Cu direction.
Hence, within a stripe model, the stripes of spin and charge are
perfectly collinear with the underlying Cu-O-Cu directions.  In
contrast, high-resolution neutron scattering measurements on stage-4
\LaOxyns~\cite{Lee99} indicate that the SDW wavevectors are rotated from
perfect alignment by $\theta_{Y} \sim 3.3^{\circ}$.  This observation of
rectangular-shifted peak positions indicates that the magnetic
ordering need not lie along high-symmetry directions of the underlying
CuO$_2$ plane.  The shift of SDW peaks, or the {\em Y-shift}, implies that
within a stripe scenario the stripes are slanted.  This presents a challenge to
existing theoretical descriptions of the magnetism in high-\tc
cuprates.  It is thus crucial to re-examine the geometry of the
elastic magnetic peaks in the \LaTwelve material.

In this paper, we report high-resolution neutron diffraction studies
of the incommensurate elastic peaks in a single crystal of
\LaTwelvens.  We have studied the same crystal ($6\phi \times 35$~mm)
previously examined in Ref.~\onlinecite{Kimura99}.  The magnetic
susceptibility shows bulk superconductivity with $T_{c}=31.5 (26.5)$~K
for the onset (midpoint) temperature of the transition.  The lattice
constants at room temperature were determined by powder x-ray
diffraction on a crushed piece of the single crystal.  In the high
temperature tetragonal (HTT) phase, the lattice constants are
$a=3.774$~\AA\ and $c=13.229$~\AA~at $T=298$~K.  Upon cooling, the material
undergoes a structural transition to the low temperature orthorhombic
(LTO) phase.  The structural transition temperature $T_{s1}$ was
determined by neutron diffraction measurements of the structural
superlattice reflection.  In this crystal, we find $T_{s1}=240$~K
which implies a Sr concentration of $x=0.12\pm0.004$~\cite{Yamada98}.
Further characterization of this crystal is given
elsewhere\cite{Kimura99}.

Neutron scattering experiments were performed using the TOPAN
triple-axis spectrometer in JRR-3M at the Japanese Atomic Energy
Research Institute.  The incident neutron energy was fixed at 14.7~meV
with pyrolytic graphite (PG) filters inserted before and after the
sample for our elastic scattering measurements.  The $(0\ 0\ 2)$
reflection of PG crystal was used to monochromatize and analyze the
neutrons.  We chose very tight horizontal collimations of
$15'-10'-{\rm Sample}-10'-{\rm Blank}$ for the dual purposes of
achieving high $q$-space resolution and improving the
signal-to-background ratio.  In this paper, we use reciprocal lattice
units for the tetragonal I4/{\it mmm} structure, even though the
crystal is orthorhombic at low temperatures.  This convention is used
in order to be consistent with our previous neutron scattering
\linebreak
\begin{figure}
\centerline{\epsfxsize=3in\epsfbox{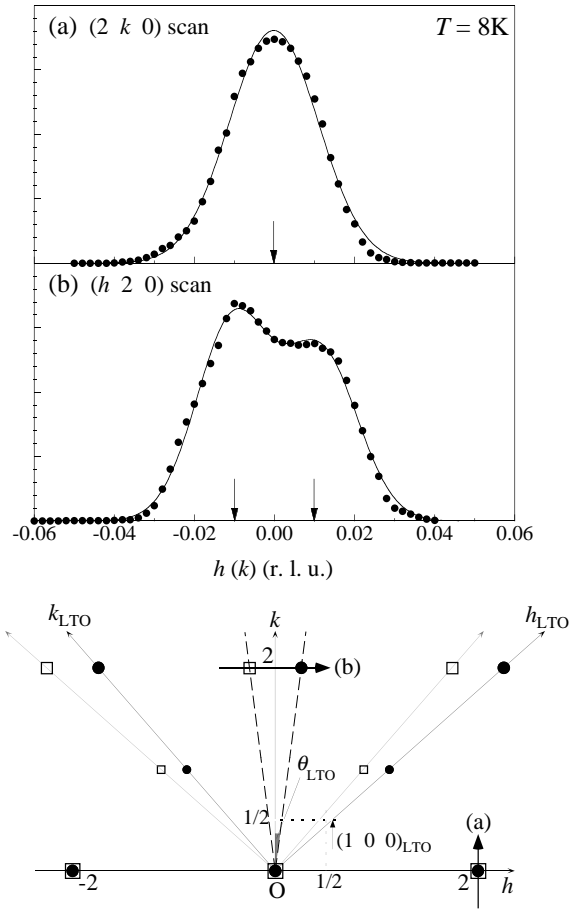}}
\vspace*{3mm}
\caption{Transverse scans through (a) the $(2\ 0\ 0)$ and (b) the
  $(0\ 2\ 0)$ nuclear Bragg peaks at $T=8$~K.  (c) A schematic diagram
  of the $(H K 0)$ reciprocal lattice depicting the two dominant twin
  domains.  Trajectories for the scans in (a) and (b) are represented
  by the arrows.}
\label{fig1}
\end{figure}
\noindent
results~\cite{Kimura99} on this crystal.  The crystal was mounted in
the $(H\ K\ 0)$ scattering zone and cooled using a closed-cycle
$^{4}$He cryostat.

We first separate the effects of structural twinning from the magnetic
scattering.  The LTO phase is characterized by a twin structure
consisting of four possible domains, which give rise to four sets of
nuclear Bragg peaks in the $(H,K,0)$ zone.  The relative populations
of these domains are sample-dependent.  Thus, we examined the peak
profiles of the nuclear Bragg peaks in the LTO phase of this crystal
in detail.  As shown in Figs.~\ref{fig1}(a) and (b), a transverse scan
through the $(2\ 0\ 0)$ Bragg peak position shows a single peak,
whereas a transverse scan through the $(0\ 2\ 0)$ position shows two
peaks.  This indicates that our crystal is primarily populated by two
structural twin domains.  These domains give rise to two sets of peaks
in reciprocal space which are simultaneously observed, as depicted in
Fig.~\ref{fig1}(c).

Now that the structural twinning has been well characterized, we turn
our focus to the elastic magnetic 
\linebreak
\begin{figure}
\centerline{\epsfxsize=3in\epsfbox{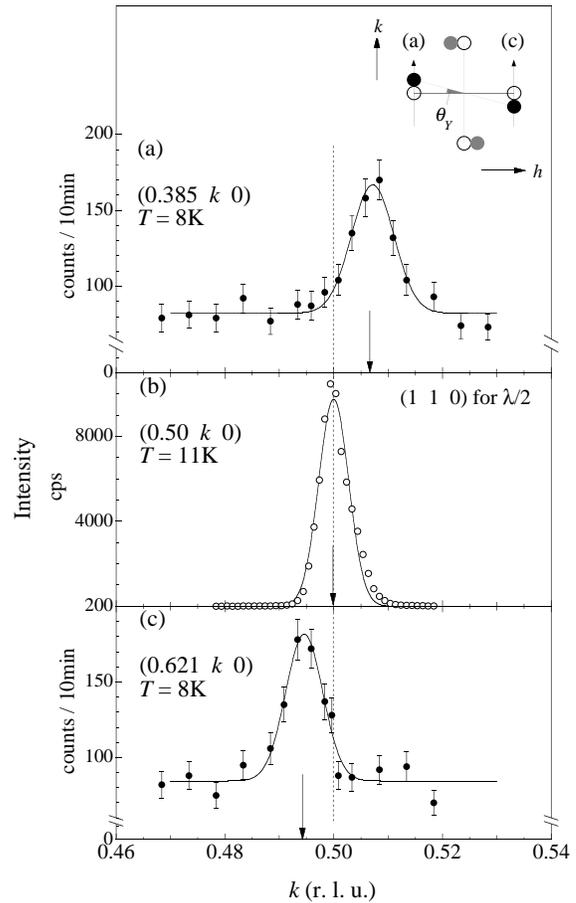}}
\vspace*{3mm}
\caption{Scans along $K$ through the elastic magnetic peaks around
  ($\frac{1}{2}\pm\varepsilon\ \frac{1}{2}\ 0$) at 8~K are shown in
  (a) and (c).  In (b), the $(1\ 1\ 0)$ nuclear Bragg peak is shown
  using $\frac{\lambda}{2}$ neutrons from the incident beam.  Scan
  trajectories for (a) and (c) are shown in the inset of (a).}
\label{fig2}
\end{figure}
\noindent 
scattering. In previous work,
Ref.~\onlinecite{Kimura99}, the authors assumed that the
incommensurate elastic magnetic peaks appear at
$(\frac{1}{2}\pm\varepsilon\ \frac{1}{2}\ 0)$ and $(\frac{1}{2}\ 
\frac{1}{2}\pm\varepsilon\ 0)$, where $\varepsilon = 0.118$.  Using
high instrumental resolution, we have reexamined the positions of the
incommensurate peaks.  We find that two of the peak positions are
given by $(\frac{1}{2}-\varepsilon\ \frac{1}{2}+\eta\ 0)$ and
$(\frac{1}{2}+\varepsilon'\ \frac{1}{2}-\eta'\ 0)$, where
$\varepsilon=0.115$, $\varepsilon'=0.121$, $\eta=0.007$ and $\eta'=0.
006$ with an error bar of $\pm0.001$.  The non-zero values of $\eta$
and $\eta'$ indicate that the peaks are shifted off of the Cu-O-Cu
axes.  This is indicated by the filled circles in Fig.~\ref{fig2}(a)
which form a rectangular quartet of incommensurate peaks (as opposed
to a square arrangement).  Alternatively, this peak-shift can be
characterized by a tilt angle $\theta_{Y} \simeq 3.0^{\circ}$ between
the incommensurate wavevector and the Cu-O-Cu direction, also depicted
in the inset of Fig.~\ref{fig2}(a).

We can show that the shifted positions of the magnetic peaks are {\it
not} an artifact due to crystal misalignment or the presence of
multiple structural twin domains.  Fig.~\ref{fig2}(b) shows a scan
along $k$ through the $(1\ 1\ 0)$ nuclear Bragg reflection using higher
order $\frac{\lambda}{2}$ neutrons from the incident beam.  This peak
is observed exactly at ($\frac{1}{2}$ $\frac{1}{2}$ 0), as expected.
This rules out the possibility that the peak shift of the elastic
magnetic peak is due to misalignment of the crystal.  Measurements of
various nuclear Bragg peaks, summarized in Fig.~\ref{fig1}(c),
indicate that peaks from different twin domains are at most offset
from each other by $\sim0.3^{\circ}$ ($\theta_{\rm LTO}$).  Since $\theta_{Y}\sim3.0^\circ$
is ten times larger, the orthorhombicity and/or concomitant twin
structure cannot account for the magnitude of the shift in the
magnetic peak positions.  Furthermore, a similar magnitude for the
magnetic peak-shift has been observed in \LaOxyns~\cite{Lee99}, even
though the orthorhombicity is about twice as great as that in
\LaTwelvens.  Therefore, we conclude that the shift of the magnetic peaks, 
i.e., the {\em Y-shift}, is an intrinsic property of the SDW order.  While striking, the reduced
symmetry of the rectangular quartet of SDW peaks is not completely
unexpected since the orthorhombicity has already broken the fourfold
symmetry of the underlying CuO$_2$ lattice.

\begin{figure}
\centerline{\epsfxsize=3in\epsfbox{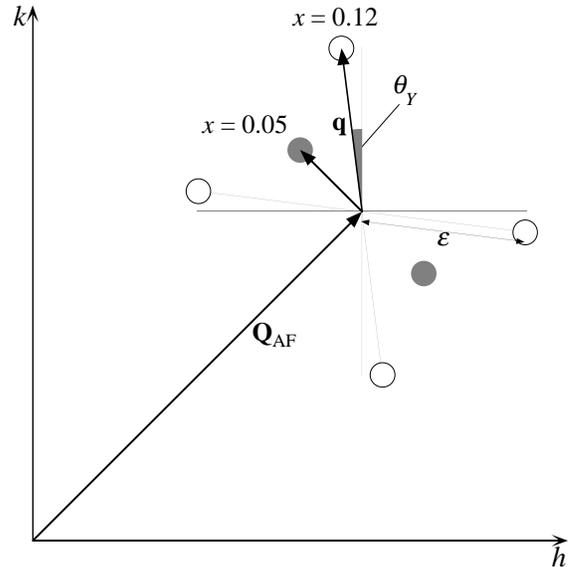}}
\vspace*{3mm}
\caption{Schematic drawing of the elastic magnetic peak positions
  in \LaSr for $x=0.05$ and 0.12.}
\label{fig3}
\end{figure}

The importance of the orthorhombicity becomes obvious by noting that
the SDW peak positions are most simply described using orthorhombic
notation.  Recall that using tetragonal notation, we found that the
incommensurabilities for the magnetic peaks, $\varepsilon$ and
$\varepsilon'$, were significantly different.  Using orthorhombic {\it
  Bmab} notation, we find that the quartet of incommensurate peaks are
precisely centered at $(1\ 0\ 0)_{ortho}$, where the low temperature
lattice constants are $a=5.3184$~\AA, $b=5.3450$~\AA, and $c=13.175$~\AA.  As shown
in Fig.~\ref{fig3}, only two parameters are required to describe the
SDW peak positions centered at $(1\ 0\ 0)_{ortho}$: a single
wavevector magnitude $\varepsilon\simeq0.118$ and a tilt angle
$\theta_{Y}\simeq3.0^\circ$.  This description for the SDW is equivalent
to that found by Lee \et\cite{Lee99} in stage-4 \LaOxy for which
$\theta_{Y}\simeq3.3^{\circ}$, a tilt angle almost identical to the
present result for \LaTwelve within the errors.  Further similarities
to the \LaOxy material are found. 
\linebreak
\noindent 
In \LaOxyns, the incommensurate
elastic peaks are centered at $(1\ 0\ l)_{ortho}$ (for even $l$) and
$(0\ 1\ l)_{ortho}$ (for odd $l$), implying the same local spin
stacking arrangement as that in the undoped \LaPure insulator.  Also,
the ordered spin direction of the SDW in \LaOxy is deduced to be along
the orthorhombic $b$-axis, again, identical to that in
La$_{2}$CuO$_{4}$\cite{Vaknin87}.  We have preliminary measurements of
the {\it L}-dependence of the elastic peaks in \LaTwelvens.  We
observe broad and weakly $q$-dependent peaks (not shown), implying
short-ranged magnetic correlations along the $c$-axis.  The result is
qualitatively consistent with that for \LaOxyns~\cite{Lee99}, implying
a similar stacking arrangement and spin direction for the ordered
spins in \LaTwelvens.  In \LaOxyns, the ordered moment had been
estimated to be $0.15\pm0.05 \mu_B$.  Comparing data taken on the same
spectrometer under identical conditions, the ordered moment of the SDW
in \LaTwelve is approximately the same as that in \LaOxyns, within the
errors.

In summary, we have established that the SDW peaks in superconducting
La$_{0.88}$Sr$_{0.12}$CuO$_{4}$ have an identical geometry, within
error, to that in \LaOxyns, namely the {\em Y-shift}.  That is, the SDW
peaks are centered around the $(1\ 0\ 0)_{ortho}$ position with a peak-shifted
rectangular arrangement.  This is in contrast to the SDW peaks in
La$_{1.6-x}$Nd$_{0.4}$Sr$_{x}$CuO$_{4}$ which retain square-symmetry
and do not show this peak-shift.  In general, the existence of short
or long-range ordered SDW peaks appears to be a common feature of the
\LaPurens-type superconductors for incommensurabilities near
$\frac{1}{8}$.  However, the {\it orthorhombic} superconductors,
\LaTwelve and stage-4 \LaOxyns, exhibit unique characteristics for the
SDW order.  At base temperature, the static spins are correlated over
very long ranges: greater than 200~\AA~in \LaTwelve \cite{Kimura99} and greater than
400~\AA~in stage-4 \LaOxyns \cite{Lee99}.  We now have established that the SDW
modulations have a reduced-symmetry peak-shift in both systems.  A
clear implication is that, in a stripe model, the stripes are slanted.
Any possible correspondence between slanted stripes and high
superconducting \tcns's requires further work.

\section*{Acknowledgments}

We acknowledge K. Machida, T. Imai, H. Fukuyama, S. Maekawa, T.
Tohyama, J. M. Tranquada, and G. Aeppli, for valuable discussions.
This work was supported in part by a Grant-In-Aid for Scientific
Research from the Japanese Ministry of Education, Science, Sports and
Culture, and by a Grant for the Promotion of Science from the Science
and Technology Agency and also supported by CREST and the US-Japan
cooperative research program on Neutron Scattering.
Work at Brookhaven National Laboratory was carried out under contact
No.  DE-AC02-98CH10886, Division of Material Science, U. S. Department
of Energy.

\

\noindent $^{*}$Present address: Research Institute for Scientific
Measurements, Tohoku University,
Katahira 2-1-1, Aoba-ku, Sendai 980-8577, Japan.\\

\noindent $^{\dag}$Present address: Institute for Materials Research,
Tohoku University, Katahira 2-1-1, Aoba-ku, Sedani 980-8577, Japan.\\

\noindent $^{\ddag}$Present address: Center for Neutron Research, NIST,
Gaithersburg, MD 20899. \\

\end{document}